\documentstyle[epsfig]{texas}
\begin{document}
\title{Testing string theory via black hole space-times}
\author{Roberto CASADIO$^{1}$ and Benjamin HARMS$^{2}$} 
\address{(1) Dipartimento di Fisica, Universit\`a di Bologna and 
\\
Istituto Nazionale di Fisica Nucleare, Sezione di Bologna,
\\
via Irnerio 46, 40126 Bologna, Italy
\\}
\address{(2) Department of Physics and Astronomy
\\
The University of Alabama
\\
Tuscaloosa, Alabama, 35487-0324, USA
\\
{\rm Email: casadio@bo.infn.it, bharms@bama.ua.edu}}
\begin{abstract}
Charged black holes, both spherically symmetric and rotating,
in the low energy limit of string theory (Einstein-Maxwell-dilaton
theory) are compared to analogous geometries in pure general
relativity.
We describe various physical differences and investigate some
experiments which can distinguish between the two theories.
In particular we discuss the gyro-magnetic ratios of rotating
black holes and the propagation of light on black hole backgrounds.
For the former we obtain an expression in the Einstein frame (EF)
which is different from the one in the String frame (SF).
This (and other results) can be used to test the stringy nature
of matter.
For a binary system consisting of a star and a rotating black hole,
we give estimates of the damping of electro-magnetic radiation coming
from the star due to the existence of a scalar component of gravity.
\end{abstract}
\section{Introduction}
It is generally accepted that super-string theory compactified down
to four space-time dimensions furnishes a description of curved
backgrounds as non-vanishing expectation values of massless string
excitations (moduli) and reproduces Einstein's general relativity
(see {\em e.g.}, \cite{polchinski} and Refs. therein).
\par
The first logical step in this derivation is thus compactification of
six extra dimensions, followed by the low energy limit, in which only
massless modes survive, and by the small coupling limit:
\par
{\centerline {\bf Superstring Theory in 9+1 dimensions}}
{\centerline {$\Downarrow$}}
{\centerline {{\bf Compactification}: $9+1\rightarrow3+1$} }
{\centerline {$\Downarrow$}}
{\centerline {{\bf Low energy}: massive modes decouple}}
{\centerline {$\Downarrow$}} 
{\centerline {{\bf Small coupling}: $\lambda_s\ll 1$}}
{\centerline {$\Downarrow$}}
{\centerline {\bf Sigma Model in 3+1 Curved Space-time:}}
\begin{equation}
S_\sigma={1\over 2\,\lambda_s^2}\,\int d^4x\,\left[h^{\alpha\beta}\,
G_{ij}\,\partial_\alpha x^i\,\partial_\beta x^j
+\epsilon^{\alpha\beta}\,B_{ij}\,\partial_\alpha x^i\,\partial_\beta x^j
+\ldots\right]
\label{s}
\end{equation}
where
$G_{ij}$ is the metric field ($i,j=0,\ldots,3$),
$B_{ij}$ an antisymmetric field (the axion potential),
$\lambda_s$ the string length, $h_{\alpha\beta}$ the world-sheet
Minkowski metric tensor,
$\epsilon^{\alpha\beta}$ the Levi-Civita symbol in 2
dimensions ($\ldots$ stand for other fields).
\par
Then one assumes the above steps do not destroy conformal symmetry
on the world-sheet and obtains a set of constraints on the fields
in the action.
Those constraints can be derived as equations of motion from an
effective action in which the string degrees of freedom $x^i$ have
formally disappeared and a new (scalar) field $\phi$ is required:
\par
{\centerline {$\Downarrow$}}
{\centerline {\bf Conformal Invariance of $S_\sigma$ on the
World-sheet}}
{\centerline {$\Downarrow$}}
{\centerline {\bf Renormalization Group Equations for the Fields
$G_{ij}$, $B_{ij}$, \ldots}}
{\centerline {$\Updownarrow$}}
{\centerline {\bf Effective Action in the String Frame ($B_{ij}=0$):}}
\begin{equation}
S_{SF}={1\over 2}\,\int d^4 x\,\sqrt{-G}\,e^{-\phi}\left[
{1\over \lambda_s^2}\,\left(R+G^{ij}\,\nabla_i\phi\,\nabla_j\phi\right)
-{1\over\alpha^2}\,e^{(1-a)\,\phi}\,F^2\right]
\label{s_s}
\end{equation}
where we have now included the action for an electro-magnetic field.
\par
The field equations are ($\lambda_s=\alpha=1$):
\begin{eqnarray}
&&
R_{ij}-{1\over 2}\,G_{ij}\,R+{1\over 2}\,G_{ij}\,\left(\nabla\phi\right)^2
-G_{ij}\,\nabla^2\phi+\nabla_i\nabla_j\phi
-2\,e^{(1-a)\,\phi}\,T_{ij}^{EM}=0
\nonumber \\
&&
\nabla^2\phi-\left(\nabla\phi\right)^2+a\,e^{(1-a)\,\phi}\,F^2=0
\nonumber \\
&&
\nabla_i\left(e^{-a\,\phi}\,F^{ij}\right)=0
\label{eq_s}
\end{eqnarray}
where $R$ is the scalar curvature of the metric $G_{ij}$,
$\nabla$ the covariant derivative with respect to $G_{ij}$,
$\phi$ the dilaton field, $\alpha$ the electro-magnetic coupling constant
and $a$ the dilaton coupling constant ($a=1$ for string theory).
The electro-magnetic energy-momentum tensor is
\begin{eqnarray}
T^{EM}_{ij} = F_{ik}\,F^k_j-{1\over{4}}\,G_{ij}\,F^2
\ .
\label{t_em}
\end{eqnarray}
\par\noindent
The name of this picture is justified by the fact that the
uncompactified degrees of freedom of the string $X^i$ move along
geodesics of the metric $G_{ij}$.
\par
The latter action can be further modified by rescaling the metric
\par
{\centerline {$\Downarrow$}}
{\centerline {\bf Conformal Transformation:}}
\begin{equation}
G_{ij}=e^{\phi-\phi_0}\,g_{ij}
\ ,
\label{conf}
\end{equation}
{\centerline {$\Downarrow$}}
{\centerline {\bf Effective Action in the Einstein Frame:}}
\begin{equation}
S_{EF}={1\over 2}\int d^4x \,\sqrt{-g}\,\left[
{1\over\ell_p^2}\,\left(R
-{1\over 2}\,g^{ij}\,\nabla_i\phi\,\nabla_j\phi\right)
-{1\over\alpha^2}\,e^{-a\,\phi}\,F^2\right]
\label{s_e}
\end{equation}
The new field equations then read ($\ell_p=\alpha=1$):
\begin{eqnarray}
&&
R_{ij} = {1\over{2}}\,\nabla_i\phi\,\nabla_j\phi + 2\,
e^{-a\,\phi}\,T^{EM}_{ij}
\nonumber \\
&&
\nabla^2\phi+a\,e^{-a\,\phi}\,F^2=0
\nonumber \\
&&
\nabla_i\left(e^{-a\,\phi}\,F^{ij}\right) = 0
\ ,
\label{eq_e}
\end{eqnarray}
where $R$ is the curvature of the metric $g_{ij}$,
$\nabla$ the covariant derivative with respect to
$g_{ij}$, $\ell_p^2=e^{\phi_0}\,\lambda_s^2$ the Planck length,
$\phi_0$ a constant.
\par
The dilaton is unchanged and the physical (covariant) components
of the electro-magnetic field are the same in both frames.
However, the uncompactified degrees of freedom of the string do not
move along geodesics of $g_{ij}$ and the scalar curvatures
differ because of the dilaton:
\begin{eqnarray}
&&
R_{(G)}=2\,\left(\nabla\phi\right)^2-3\,\nabla^2\phi
\nonumber \\
&&
R_{(g)}={1\over 2}\,\left(\nabla\phi\right)^2
\ .
\label{R}
\end{eqnarray}
\par
Which frame is more suitable as a description of the present state of
our Universe is an open question which will eventually be settled by
experiment.
The issue of conformal transformations in theories of gravity is
well known (see, {\em e.g.}, the extensive review \cite{faraoni})
and was first raised in the context of the low energy string theory
in Ref.~\cite{dick}.
Although it can be proven that the two frames are dynamically
equivalent (the conformal transformation (\ref{conf}) is canonical
\cite{garay}), it is clear that (at most) one of the metrics
involved can be used to compute the distances and related quantities
which are actually measured in the experiments.
A common view is that strings follow geodesics of $G_{ij}$, while
``ordinary'' particles are expected to follow geodesics of $g_{ij}$.
If real particles are made of strings, a contradiction arises
because the two kinds of trajectories do not coincide in general, nor
are they related by a change of coordinates.
A possible way out of this paradox is that in one frame the
corresponding metric gives distances with respect to a fixed
reference length (which one might take to be the Compton wave-length
of massive matter fields) and a fixed interval of time ({\em e.g.},
the inverse of the frequency of some basic nuclear process),
while in the other such reference length and time interval are locally
deformed due to the extra force given by the dilaton.
Hence, the tests described below are meant to unveil the nature
of ``ordinary'' matter: if matter retains stringy aspects and the
basic length and time units at our disposal are truly constant,
one should find the values computed in SF;
on the other hand, if the EF turns out to be a good framework,
then one could infer that ordinary matter is subject to an extra
force or perhaps question the physical relevance of string theory.
\par
This issue has already been extensively discussed in the framework
of scalar-tensor theories of gravity and observable consequences have
been deduced mainly in cosmology
\cite{faraoni}.
Because of the direct coupling between the dilaton and matter
(in our case the electro-magnetic field), both actions in
Eqs.~(\ref{s_s}) and (\ref{s_e}) fail to be of the Brans-Dicke type,
thus the equivalence principle does not hold in general
(it can be reinstated in places of the Universe where
the dilaton becomes massive due to higher order corrections in
$\lambda_s$ \cite{polchinski}).
One sees the equivalence principle is violated whenever the gradient of
the dilaton field is not negligible and there are at present strong
constraints from observation on the magnitude of such violations.
However, these constraints might be ineffective provided the violations
occurred far in the past, {\em e.g.} in the early stages of the Universe,
or take place in regions of space which have not been tested directly,
{\em e.g.} near black hole (horizons), the latter being regarded as
excitations of extended objects \cite{hl1,hl2,chl1,hl3,hl4,hl5,hl6,hl7}.
%
%
\subsection{Dilatonic black holes}
Here we list some of the known solutions (either exact or in some
approximation)
of the field equations (\ref{eq_e}) in the Einstein frame for
$a\not=0$ which can be used to describe black holes with ADM mass
$M$ and are parameterized by
the values of the electric charge $Q$ and the angular momentum $J$:
\begin{description}
\item[I)] $Q=J=0$: {\bf Janis-Newman-Winicour} (exact \cite{jnw}).
\par\noindent
It represents the geometry outside a spherically symmetric,
electrically neutral source.
It contains a central naked singularity (no horizons).
\item[II)] $Q\not=0,J=0$: {\bf Reissner-Nordstr\"om dilatonic (RND)}
(exact \cite{gm,hs}).
\par\noindent
It represents the geometry outside a spherically symmetric,
electrically charged source and reduces to the Reissner-Nordstr\"om
(RN) metric for $a=0$.
\item[III)] $Q,J\not=0$:
\begin{description}
\item[i)] $a=\sqrt{3}$: {\bf Kaluza-Klein} (exact \cite{gm,hh,shira}).
\par\noindent
It coincides with a Kaluza-Klein 5 dimensional model compactified to
4 dimensions.
\item[ii)] $J/M^2\ll1$: {\bf Kerr-Newman dilatonic} (approximate
\cite{hh,shira,cko})
\par\noindent
It represents the geometry outside an axially symmetric, slowly rotating
electrically charged source and reduces to the RND metric for $J=0$
and to the Kerr-Newman (KN) metric for $a=0$.
\item[iii)] $Q/M\ll1$: {\bf  Kerr-Newman dilatonic (KND)} (approximate
\cite{knd})
\par\noindent
The geometry generated by the same kind of source as {\bf ii)}, with small
electric charge but arbitrary angular momentum.
\end{description}
\end{description}
The above solutions are mapped into solutions in the SF by
Eq.~(\ref{conf}).
It should be emphasized that the corresponding static dilaton field
falls off (to a constant value which can always be set to zero) far
from the central singularity, thus making the two frames coincide far
away from the horizon.
\section{$Q\not=0$, $J=0$: RND black holes}
The RND metric represents spherically symmetric, electrically charged
black holes for $M/Q>\sqrt{1-a^2}$, where $a$ is the dilaton coupling.
\subsection{Einstein Frame}
In EF the line element is given by \cite{gm,hs}
\begin{eqnarray}
\left.ds^2\right]_{EF}=-e^{2\,\Phi}\,dt^2+e^{2\,\Lambda}\,dr^2
+R^2\,d\Omega_2^2
\ ,
\label{rnd_ef}
\end{eqnarray}
where $d\Omega_2^2=d\theta^2+\sin^2\theta\,d\varphi^2$ and
\begin{eqnarray}
&&e^{2\,\Phi}=e^{-2\,\Lambda}
=\left(1-{r_+\over r}\right)\,
\left(1-{r_-\over r}\right)^{1-a^2\over 1+a^2}
\nonumber \\
&&R^2=r^2\,\left(1-{r_-\over r}\right)^{2\,a^2\over 1+a^2}
\ .
\end{eqnarray}
There are three singularities, an essential singularity at
$r=0$ and two coordinate singularities at
\begin{eqnarray}
&&r_+=M+\sqrt{M^2-(1-a^2)\,Q^2}
\nonumber \\
&&r_-=(1+a^2)\,{Q^2\over r_+}
\ .
\end{eqnarray}
The value $r_-$ is a weak singularity for $a\not=0$, while $r_+$ is
an horizon.
The black hole has also a static electro-magnetic field
\begin{eqnarray}
&&F_{tr}={Q\over r^2}
\ ,
\label{E}
\end{eqnarray}
and a static dilaton field
\begin{eqnarray}
e^{-\phi}=
\left(1-{r_-\over r}\right)^{-{2\,a\over 1+a^2}}
\ .
\label{phi_rnd}
\end{eqnarray}
By taking the large $r$ expansion of the metric and electro-magnetic
field,
one sees that $M$ and $Q$ represent the physical (ADM) mass and charge
of the black hole.
\subsection{Newtonian approximation}
From $e^{\phi}\sim G_N$, the total force acting on a test mass
(of constant value) $m$ is the sum of the force $F_\phi$ due to the spatial
dependence of $G_N$ and the Newtonian contribution $F_N$,
\begin{equation}
F_{tot}\sim-\partial_r\left(G_N\,{M\,m\over r}\right)
=F_\phi+G_N\,{M\,m\over r^2}
\ .
\end{equation}
We observe $F_\phi$ becomes of the same order as $F_N$ at
\begin{equation}
r\sim {1+3\,a^2\over 1+a^2}\,r_-
\ .
\end{equation}
If one wishes to perform a measurement with the precision of one part over
$10^N$, one has to go closer than $r_c\sim 10^N\,r_-$ to the black hole
centre in order to test any violation of the equivalence principle.
Since $r_c>r_+$, this gives the following estimate for the smallest
charge-to-mass ratio that the black hole must possess in order to test
any deviation:
\begin{equation}
{Q\over M}>10^{-N/2}
\ .
\end{equation}
For a solar mass black hole and $N\sim 10$ this means a charge
of about $10^{34}$ electron charges or $10^{15}$ C.
For a Planck mass black hole with one electron
charge the ratio $Q/M\sim 0.1$ and one needs $N\sim 2$.
\subsection{String Frame}
The SF metric is given by \cite{knd_es}
\begin{eqnarray}
\left.ds^2\right]_{SF}&=&-\left(1-{r_+\over r}\right)\,
\left(1-{r_-\over r}\right)^{1+2\,a-a^2\over 1+a^2}\,dt^2
\nonumber \\
&&+\left(1-{r_+\over r}\right)^{-1}\,
\left(1-{r_-\over r}\right)^{a^2+2\,a-1\over 1+a^2}\,dr^2
\nonumber \\
&&+r^2\,\left(1-{r_-\over r}\right)^{2\,a\,(1+a)\over 1+a^2}\,d\Omega_2^2
\ .
\label{rnd_sf}
\end{eqnarray}
A major consequence of the conformal rescaling is that the physical
(ADM) mass
of the black hole is shifted according to
\begin{eqnarray}
\left.M_{phys}\right]_{SF}&=&M+{a\,Q^2\over r_+}
=M\,\left(1+{a\,Q^2\over 2\,M^2}\right)
+{\cal O}\left({Q^4\over M^4}\right)
\ .
\label{m_rnd}
\end{eqnarray}
\subsection{First test: determination of the gravitational mass}
\label{t1}
According to Eq.~(\ref{m_rnd}), the physical (ADM) mass is different
in the two frames.
A possible method for discriminating between the two frames is
to compare the experimental value of $M_{phys}$ with the one computed
from the knowledge of $Q$ and $r_+$.
\par
One observes that:
\par\noindent
{\it 1)} the electric field $F_{tr}$ can be measured
by comparing the acceleration of a charged test particle to the
acceleration of a neutral particle of equal mass;
\par\noindent
{\it 2)} the value of $M_{phys}$ is obtained directly from the
acceleration of a neutral particle at large distance;
\par\noindent
{\it 3)} the radius $r_+$ can be estimated by inferring the largest
distance from which light can escape or by determining the inner edge
of the accreting disk.
\par
The electro-magnetic field is conformally invariant and $F_{tr}$
is given by Eq.~(\ref{E}) in both frames.
Thus step {\it 1)} allows the computation of $Q$ and the insertion of
$Q$ into the definition
of $r_+$ which, together with the measured value of $r_+$
from {\it 3)}, gives $M$.
If $M$ is equal to $M_{phys}$ from {\it 2)}, then EF is the physical
picture and one might question the stringy origin of the action $S_{g}$;
in case they are not equal, SF is the physical picture and (\ref{m_rnd})
can be used to estimate $a$.
\subsection{Evaporation}
The time dependence of the mass of the black hole which emits Hawking
quanta depends on the frame.
The total energy of the system is equal to $M\equiv M_{phys}$ and
constant,
therefore the micro-canonical ensemble must be implemented
\cite{knd_sm,micro}.
\par
The surface area of the outer horizon is given by
\begin{eqnarray}
&&\left.{\cal A}\right]_{EF}=4\,\pi\,r_+^{2\over 1+a^2}\,
\left(r_+-r_-\right)^{2\,a^2\over 1+a^2}
\nonumber \\
&&\left.{\cal A}\right]_{SF}=4\,\pi\,r_+^{2\,(1-a)\over 1+a^2}\,
\left(r_+-r_-\right)^{2\,a\,(1+a)\over 1+a^2}
\ ,
\end{eqnarray}
therefore the ratio
\begin{equation}
{\left.{\cal A}\right]_{EF}\over\left.{\cal A}\right]_{SF}}
=\left({r_+-r_-\over r_+}\right)^{2\,a\over 1+a^2}<1
\ \ (>1)\ \ \ \
{\rm for}\ \ a>0\ \ (a<0)
\end{equation}
\par
\begin{figure}
\centering
\epsfig{figure=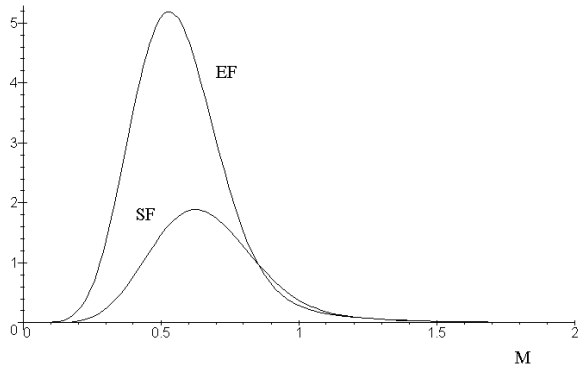,width=9cm} 
\caption{Energy emitted by RND black holes per unit time and
ratio $Q/M$ fixed in the two frames.
The mass is in units of the Planck mass.
The vertical scale is arbitrary.}
\label{a}
\end{figure}
\begin{figure}
\centering
\epsfig{figure=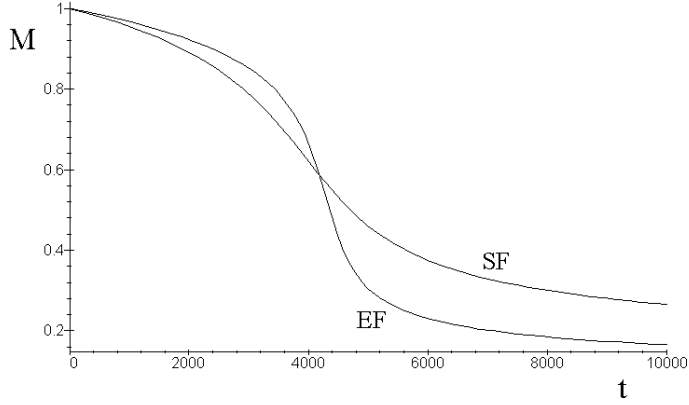,width=9cm}   
\caption{Time evolution of the mass of RND black holes with
$M(0)=M_p$ in the two frames. The time scale is arbitrary.}
\label{b}
\end{figure}

We then assume the internal degeneracy of the black hole is given
by the area law
\begin{equation}
\Omega\sim e^{{\cal A}/4}
\simeq e^{\pi\,M^2\,\left(1-f\,x^2\right)}
\end{equation}
where we have taken $a=1$, $x\equiv Q/M$ small and constant and
$f=1/2$ in EF (3/2 in SF).
Thus the micro-canonical occupation number density of the Hawking
radiation is
\begin{eqnarray}
n(\omega)&=&\sum\limits_{l=1}^{M/\omega}
{\Omega(M-l\,\omega)\over\Omega(M)}
\nonumber \\
&\sim&\sum\limits_{l=1}^{M/\omega}
\left[e^{4\,\pi\,l^2\,\omega^2-8\,\pi\,M\,l\,\omega}\right]^{(1-f\,x^2)}
\ .
\label{n_f}
\end{eqnarray}
One can now estimate the energy emitted by the black hole per unit
time as
\begin{equation}
{dM\over d\tau}\sim
-{\cal A}\,\int d\omega\,\omega^3\,\Gamma(\omega)\,n_f(\omega)
\ .
\label{dmdt}
\end{equation}
For $\Gamma=1$ and $x=1/2$ in Eq.~(\ref{dmdt}) the result is given in
Fig.~\ref{a} and \ref{b}.
For large values of $M$ the emission is approximately thermal at the
Hawking temperature $T=(1+f\,x^2)/8\,\pi\,M$ and more intense in SF,
thus leading to a faster decay.
The intensity in EF overcomes the intensity in SF for values
around the Planck mass $M_p\equiv\sqrt{\hbar\,c/G_N}$ and smaller.
Both emissions reach a maximum and then vanish for zero mass, a feature
which is a direct consequence of the use of the micro-canonical approach
(energy conservation).
\section{$J\not=0$, $Q/M\ll1$: KND black holes}
The KND metric represents rotating, electrically charged black holes
for $M^2-Q^2-\alpha^2\ge 0$ and is a more realistic candidate for the
description of astrophysical black holes which are thought to be spinning
rapidly.
\subsection{Einstein Frame}
The line element in EF \cite{knd,kndw}
\begin{equation}
\left.g_{ij}\right]_{EF}=g_{ij}^{KN}+{\cal O}\left({Q^4\over M^4}\right)
\ ,
\end{equation}
where $g_{ij}^{KN}$ is the Kerr-Newman (KN) metric ($\alpha\equiv J/M$) is,
\begin{eqnarray}
ds^2_{KN} &=&
-\sqrt{\Delta}\,\sin\theta\,\left[
\chi\,d\varphi^2-{1\over\chi}\,(dt-\omega\,d\varphi)^2\right]
\nonumber \\
&&+\rho^2\,\left[{(dr)^2\over\Delta}+(d\theta)^2\right]
\ ,
\label{g_ij}
\end{eqnarray}
with
\begin{eqnarray}
\chi&=&{\sqrt{\Delta}\,\sin\theta\over\Psi}
\nonumber \\
\Delta&=&r^2-2\,M\,r+\alpha^2+Q^2
\nonumber \\
\rho^2&=&r^2+\alpha^2\,\cos^2\theta
\nonumber \\
\Psi&=&-{\Delta-\alpha^2\,\sin^2\theta\over\rho^2}
\nonumber \\
\omega&=&-\alpha\,\sin^2\theta\,[1+\Psi^{-1}]
\ .
\label{KN-metric}
\end{eqnarray}
Thus the geodesic motions of neutral particles are unaffected by the
presence of a static dilaton field (up to order $Q^3/M^3$) and
the causal structure is not changed by the dilaton to that order.
There are two horizons at
\begin{equation}
r_\pm=M \pm\sqrt{M^2-\alpha^2- Q^2}
\ .
\end{equation}
The static dilaton field is
\begin{eqnarray}
\phi=-a\,{r\over\rho^2}\,{Q^2\over M}
\ .
\label{p0}
\end{eqnarray}
One can also compute the corresponding electric and magnetic
field potentials \cite{knd},
\begin{eqnarray}
A&=&Q\,{r\over\rho^2}\,\left[
1-\left({1\over2\, r}+{r\over\rho^2}\right)\,
{a^2\,Q^2\over 3\,M}\right]
\nonumber \\
B&=&-Q\,\alpha\,{\cos\theta\over\rho^2}\,
\left[1-\left({1\over 2\,M}-{r\over\rho^2}\right)\,
{a^2\,Q^2\over3\,M}\right]
\ .
\end{eqnarray}
Terms proportional to $Q^2$ inside the brackets above are corrections
with respect to the KN potentials, which becomes more apparent if one
writes the electric and magnetic fields for large $r$,
\begin{eqnarray}
{\cal E}_{\hat{r}} &\approx& {Q\over{r^2}}
\nonumber \\
{\cal E}_{\hat{\theta}} &\approx&
-{2\,\alpha^2\, Q\over{r^4}}\,\sin\theta\,\cos\theta
\nonumber \\
{\cal B}_{\hat{r}} &\approx&
{2\,\alpha\, Q\over{r^3}}\,\cos\theta
\,\left[1 - {a^2\,Q^2\over{6\,M^2}}\right]
\nonumber \\
{\cal B}_{\hat{\theta}} &\approx&
{\alpha\, Q\over{r^3}}\,
\sin\theta\left[1 - {a^2\,Q^2\over{6\,M^2}}\right]
\equiv{\mu_{phys}\over r^3}\,\sin\theta
\ .
\label{E_knd}
\end{eqnarray}
One thus recognizes that the asymptotic electric field is the same
as in KN,
however the intensity of the asymptotic magnetic field is lower.
\subsection{String Frame}
In SF the metric is \cite{knd_es}
\begin{eqnarray}
\left.G_{ij}\right]_{SF}
=g_{ij}^{KN}\,\left(1-a\,{r\over\rho^2}\,{Q^2\over M}\right)
+{\cal O}\left({Q^4\over M^4}\right)
\ ,
\label{G}
\end{eqnarray}
where $J_{phys}$ and $\mu_{phys}$ are not changed by the conformal
rescaling
but the ADM mass is shifted to
\begin{equation}
\left[M_{phys}\right]_{SF}=\left[M_{phys}\right]_{EF}\,
\left(1+{a\,Q^2\over 2\,M^2}\right)
\ .
\end{equation}
Therefore the same experimental test as described in section~\ref{t1}
works for KND.
\subsection{Second test: the gyro-magnetic ratio}
Rotating charged KN black holes have a gyro-magnetic ratio
\cite{strau}
\begin{eqnarray}
g=2\,{\mu_{phys}\,M_{phys}\over Q_{phys}\,J_{phys}}=2
\ .
\end{eqnarray}
However KND black holes posses an anomalous, frame dependent,
gyro-magnetic ratio \cite{knd,knd_es}
\begin{eqnarray}
&&[g]_{EF}\simeq 2\,\left[1-{a^2\,Q^2\over 6\, M^2}\right]<2
\label{gyro_ef}
\\
&&[g]_{SF}\simeq 2\,\left[1+{a\,Q^2\over 2\,M^2}\,\left(1-{a\over 3}\right)
\right]
\left\{\begin{array}{ll}
<2 &\ \ \ \ a<0\ ,\ \ a>3
\\
=2 &\ \ \ \ a=3
\\
>2 &\ \ \ \ 0<a<3
\end{array}\right.
\label{gyro_sf}
\end{eqnarray}
The latter case provides another way of testing which picture is the
physical one.
In fact, since $[g]_{EF}$ can be at most equal to $2$, the measurement
of a value greater than $2$ for the gyro-magnetic ratio of a black hole
would prove that physics has to be described in SF (we remark that from
string theory $a=1$).
On the other hand, the measurement of any value smaller than $2$,
although crucial for proving the existence of static dilaton field,
would not suffice for discriminating between EF and SF, unless an
independent way of measuring $a$ along with the mass and charge of the
black hole can be found.
\section{More tests: light propagation}
Since the relation between SF and EF is given by a conformal
transformation of the metric, eikonal paths followed by null rays
are the same in the two frames, as are the deflection angles of light
scattered by the black hole.
In particular, for KND this means that, to lowest order in $(Q/M)^2$,
light rays are not affected by the dilaton \cite{kndw2}.
However, one can use the fact that proper distances and times of flight
depend on the frame.
\subsection{Reverberation}
\begin{figure}
\centering
\epsfig{figure=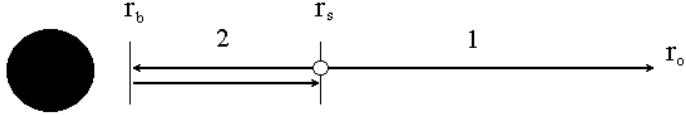,width=9cm}   
\caption{Simple model of {\em reverberation}.}
\label{c}
\end{figure}

A way of detecting time delays is displayed in Fig.~\ref{c}
\cite{knd_es}.
A light source is at $r=r_s$ and emits both
towards the observer placed at $r_o$ (ray 1) and towards the black hole
(ray 2).
The latter ray then bounces back at $r_b$ and reaches the observer
with a delay with respect to ray 1 given by twice the time it takes to
go from the source to $r_b$.
In EF this delay is given by (again assuming RND with $a=1$)
\begin{eqnarray}
\left.\tau\right]_{EF}\sim 2\,\left[r_s-r_b
+r_+\,\ln\left({r_s-r_+\over r_b-r_+}\right)\right]
\ ,
\end{eqnarray}
while in SF one has
\begin{eqnarray}
\left.\tau\right]_{SF}\sim\left.\tau\right]_{EF}
-2\,r_-\,\ln\left({r_s-r_+\over r_b-r_+}\right)
\ .
\end{eqnarray}
The difference depends only on the positions of the source and of
the reflection point (presumably inside the accreting disk).
\subsection{Red-shift}
The difference between the metrics in the two frames also affects
the red-shift $z$ of waves emitted at $r_s$ \cite{knd_es}.
For instance, in RND with $a=1$ one has
\begin{eqnarray}
&&\left.z\right]_{EF}=-{r_+\over r_s}=-{2\,M\over r_s}
\nonumber  \\
&&\left.z\right]_{SF}=\left.z\right]_{EF}-{r_-\over r_s}+
{r_-\,r_+\over r_s^2}=\left.z\right]_{EF}
-{Q^2\over r_s}\,\left({1\over M}-{2\over r_s}\right)
\ .
\end{eqnarray}
Since typically $r_s>2\,M$, $\left.z\right]_{EF}>\left.z\right]_{SF}$.
\subsection{Linear waves in RND}
Linear perturbation theory applied to the KND solution gives a set of
coupled wave equations for the electro-magnetic, dilaton and gravitational
fields \cite{kndw,kndw2}.
Those equations can be conveniently analyzed by expanding in $Q/M$.
The processes at lowest order are:
\begin{description}
\item[1)]
EM waves $F^{(1)}$ interact with static EM background $F^{(0)}$ and
produce dilaton waves $\phi^{(1)}$
\item[2)]
EM waves $F^{(1)}$ interact with static EM background $F^{(0)}$ and
produce gravitational waves $G^{(1)}$
\item[3)]
Dilaton waves $\phi^{(1)}$ interact with static EM background $F^{(0)}$ and
produce EM waves $F^{(1)}$
\end{description}
\par
For EM waves, although at leading order the eikonal trajectories
are the same in both frames, the intensity of the produced waves
$F^{(1)}$ in {\bf 3)} is different because of the different metric
backgrounds $G^{(0)}$.
From (\ref{conf}) one can estimate the intensity of electro-magnetic
radiation produced in scattering events involving other fields in the
model according to $I\sim |\vec E|^2+|\vec B|^2\sim F^2$, that is
\begin{eqnarray}
[I]_{SF}\sim e^{-2\,\phi}\,\left[F^2\right]_{EF}\sim e^{-2\,\phi}\,
[I]_{EF}
\ .
\end{eqnarray}
This implies $[I]_{SF}\sim [I]_{EF}\,\left(1+{\cal O}(r^{-1})\right)$
and the difference is appreciable when the scatterings occur near
the horizon.
\begin{figure}
\centering
\epsfig{figure=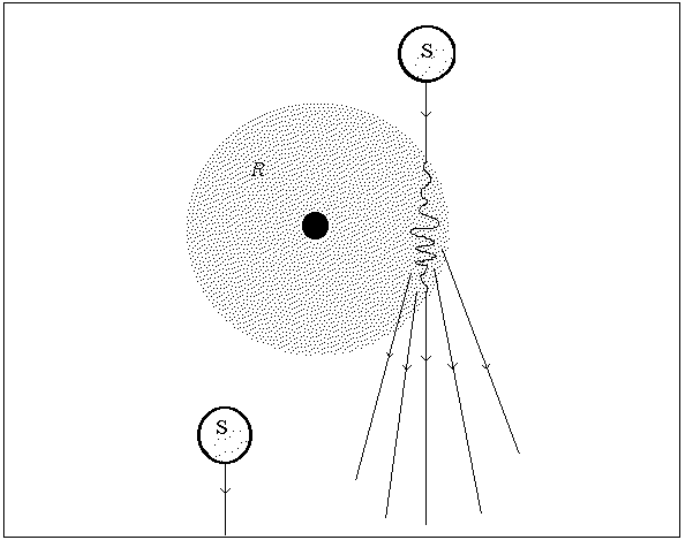,width=9cm} 
\caption{Binary system made of a star and a RND black hole.}
\label{fig4}
\end{figure}
\subsection{Star spectrum in a binary system}
A case of particular interest is given by {\bf 1)} since it allows
the computation of the energy transferred from the spectrum of a star
to the dilaton field of an RND companion, Fig.~\ref{fig4} \cite{binary}.
When the star is going behind the black hole, its radiation toward
the observer passes near the horizon and stimulates dilaton waves, thus
losing energy.
The corresponding spectrum can then be compared to the unperturbed one
obtained when the star is in front of the black hole.
\section*{References}
\end{document}